\documentclass[twocolumn,showpacs,preprintnumbers,amsmath,amssymb]{revtex4}

\usepackage{graphicx}
\usepackage{dcolumn}
\usepackage{bm}

\begin{document}


\title{Determination of polarized PDFs from a QCD analysis of inclusive and
semi-inclusive Deep Inelastic Scattering data  }

\author{Elliot Leader}
\email{e.leader@imperial.ac.uk} \affiliation{Imperial College,
Prince Consort Road, London SW7 2BW, England.
}%

\author{Aleksander V. Sidorov}
\email{sidorov@theor.jinr.ru} \affiliation{ Bogoliubov Theoretical
Laboratory Joint Institute for Nuclear Research 141980 Dubna,
Russia.
}%

\author{Dimiter B. Stamenov}
\email{stamenov@inrne.bas.bg}
\affiliation{ Institute for Nuclear Research and Nuclear Energy \\
Bulgarian Academy of Sciences
Blvd. Tsarigradsko Chaussee 72, Sofia 1784, Bulgaria}%

\begin{abstract}
{\bf Abstract}

A new combined next to leading order QCD analysis of the polarized
inclusive and semi-inclusive deep inelastic lepton-hadron
scattering (DIS) data is presented. In contrast to previous
combined analyses, the $1/Q^2$ terms (kinematic - target mass
corrections, and dynamic - higher twist corrections) in the
expression for the nucleon spin structure function $g_1$ are taken
into account. The new COMPASS data are included in the analysis.
The impact of the semi-inclusive data on the polarized parton
densities (PDFs)  and on the higher twist corrections is
discussed. The new results for the PDFs are compared to our
(Leader, Sidorov, Stamenov) LSS'06 PDFs, obtained from the fit to
the inclusive DIS data alone, and to those obtained from the de
Florian, Sassot, Stratmann, and Vogelsang global analysis.

\end{abstract}

\pacs{13.60.Hb, 12.38.-t, 14.20.Dh}
\maketitle

\section{Introduction}

Experiments on polarized inclusive deep inelastic lepton-hadron
scattering (DIS), reactions of the type $l+p\rightarrow l' + X $
with both polarized lepton and hadron, because of the nonexistence
of neutrino data, can only, in principle yield information on the
sum of quark and antiquark parton densities i.e. information on
the polarized densities $\Delta u + \Delta \bar{u},~\Delta d +
\Delta \bar{d},~ \Delta s + \Delta \bar{s}~\textrm{and}~ \Delta
G$.

Information about the antiquark densities $\Delta \bar{u}  , \, \Delta
\bar{d}$ and the separate $\Delta s $ and $\Delta \bar{s}$ strange
densities thus has to be extracted
from other reactions, notably polarized semi-inclusive lepton-hadron
reactions (SIDIS) $l+p\rightarrow l' +h+X$, where $h$ is a detected
hadron in the final state, or from semi-inclusive hadronic scattering
(SIHS) like $p+p\rightarrow h + X$, involving polarized protons, and
only possible at the RHIC collider at Brookhaven National
Laboratory.

QCD analyses of polarized DIS data, at next to leading order
accuracy (NLO), have been carried out for some decades (for more
recent analyses see \cite{{groups}, {LSS07}}), but it was only in
2008 that de Florian, Sassot , Strattmann and Vogelsang (DSSV), in
a ground-breaking paper \cite{DSSV}, carried out a combined
analysis of polarized DIS, SIDIS and SIHS, at NLO accuracy.

The technical problems involved in going from an analysis of DIS to
such a combined analysis are formidable. In this paper we extend our
study of polarized DIS to a joint analysis of the world data on DIS
and SIDIS reactions.

In contrast to the situation in unpolarized DIS, a large portion
of the most accurate data on polarized DIS lie in a kinematical
region where target mass corrections (TMC) of order $M^2/Q^2$ (
whose form is exactly known), and dynamical higher twist (HT)
corrections of order $\Lambda^2_{QCD}/Q^2$ are important. We have
thus included such terms in our description of the DIS data.
However, for the SIDIS data, we do not know  the analogous results
at present, so do not include such terms. As it happens almost all
the SIDIS data we utilize are in kinematic regions where such
corrections should not be important.

Despite the fact that it has been emphasized in the literature for
more than a decade that DIS data can only, in principle, yield
information  on the sum of quark and antiquark densities, some
analyses of purely inclusive DIS continue to show results for
valence densities, under what are termed \emph{assumptions} about
the sea-quark densities $\Delta \bar{u}(x)$ and $\Delta
\bar{d}(x)$. It is important to realize that these are not really
physical assumptions, but merely conventions. In contrast, it is
important to note that although we tend to think of the strange
quark density as a sea-quark density, $\Delta s(x) + \Delta
\bar{s}(x) $ is fully determined by the purely inclusive DIS data.
This is particularly important because of the apparent
incompatibility of the polarized strange quark density  obtained
from DIS and from SIDIS data, as will be discussed in detail
later.

In this paper we present the results of our NLO QCD analysis of
polarized inclusive and semi-inclusive DIS data. Our analysis
differs from DSSV in the following respects:

~~(i) We have included new data from the COMPASS group at CERN,
which were not available in 2008.

~(ii) We are more careful in handling the kinematics and include
target mass corrections  and higher twist terms  in the DIS sector
of our analysis.

(iii) Our parametrization of the parton densities is similar to
that of DSSV, but differs in some important aspects, as will be
explained in detail in Sec. III.

\section{QCD framework for inclusive and semi-inclusive polarized DIS}

\subsection{Inclusive DIS}
One of the features of polarized DIS is that more than half of the
present data are at moderate $Q^2$ and $W^2$ ($Q^2 \sim 1-5~\rm
GeV^2,~4~\rm GeV^2 < W^2 < 10~\rm GeV^2$), or in the so-called
{\it preasymptotic} region. This is especially the case for the
very precise experiments performed at the Jefferson Laboratory.
So, in contrast to the unpolarized case this region cannot be
excluded from the analysis. As was shown in \cite{LSS_HT}, to
confront correctly the QCD predictions to the experimental data
including the preasymptotic region, the {\it nonperturbative}
higher twist (powers in $1/Q^2$) corrections to the nucleon spin
structure functions have to be taken into account too.

In QCD the spin structure function $g_1$ has the following form
for $Q^2 >> \Lambda^2$ (the nucleon target label $N$ is not
shown):
\begin{equation}
g_1(x, Q^2) = g_1(x, Q^2)_{\rm LT} + g_1(x, Q^2)_{\rm HT}~,
\label{g1QCD}
\end{equation}
where "LT" denotes the leading twist ($\tau=2$) contribution to
$g_1$, while "HT" denotes the contribution to $g_1$ arising from
QCD operators of higher twist, namely $\tau \geq 3$:
\begin{eqnarray}
g_1(x, Q^2)_{\rm LT}&=& g_1(x, Q^2)_{\rm pQCD}  + h^{\rm TMC}(x,
Q^2)/Q^2\nonumber\\
&&+ {\cal O}(M^4/Q^4)~, \label{g1LT}
\end{eqnarray}
where $g_1(x, Q^2)_{\rm pQCD}$ is the well-known (logarithmic in
$Q^2$) NLO perturbative QCD contribution
\begin{eqnarray}
g_1(x,Q^2)_{\rm pQCD}&=&\frac{1}{2}\sum _{q} ^{n_f}e_{q}^2
[(\Delta q +\Delta\bar{q})\otimes (1 +
\frac{\alpha_s(Q^2)}{2\pi}\delta C_q)
\nonumber\\
&&+\frac{\alpha_s(Q^2)}{2\pi}\Delta G\otimes \frac{\delta C_G}
{n_f}], \label{g1partons}
\end{eqnarray}
and $h^{\rm TMC}(x, Q^2)$ are the exactly calculable kinematic
target mass corrections \cite{TMC}, which, being purely kinematic,
effectively belong to the LT term. In Eq. (\ref{g1partons}),
$\Delta q(x,Q^2), \Delta\bar{q}(x,Q^2)$ and $\Delta G(x,Q^2)$ are
quark, antiquark and gluon polarized densities in the proton,
which evolve in $Q^2$ according to the spin-dependent NLO DGLAP
equations. $\delta C(x)_{q,G}$ are the NLO spin-dependent Wilson
coefficient functions and the symbol $\otimes$ denotes the usual
convolution in Bjorken $x$ space. $n_f$ is the number of active
flavors ($n_f=3$ in our analysis). In addition to the LT
contribution, the dynamical higher twist effects
\begin{equation}
g_1(x, Q^2)_{\rm HT}= h(x, Q^2)/Q^2 + {\cal O}(\Lambda^4/Q^4)~,
\label{HTQCD}
\end{equation}
must be taken into account at low $Q^2$. The latter are
nonperturbative effects and cannot be calculated in a model
independent way. That is why we prefer to extract them directly
from the experimental data. The method used to extract
simultaneously the polarized parton densities and higher twist
corrections to $g_1$ from data on $g_1/F_1$ and $A_1(\approx
g_1/F_1)$, is described in \cite{LSS_HT}. Accordingly the
$g_1/F_1$ data have been fitted using the experimental data for
the unpolarized structure function $F_1(x,Q^2)$:
\begin{equation}
\left[\frac{g_1(x,Q^2)}{F_1(x, Q^2)}\right]_{exp}~\Leftrightarrow~
\frac{{g_1(x,Q^2)_{\rm LT}+h(x)/Q^2}}{F_1(x,Q^2)_{exp}}~.
\label{g1F2Rht}
\end{equation}

As usual, $F_1$ is replaced by its expression in terms of the
usually extracted from unpolarized DIS experiments $F_2$ and $R$.
As in our previous analyses, the  phenomenological
parametrizations of the experimental data for $F_2(x,Q^2)$
\cite{NMC} and the ratio $R(x,Q^2)$ of the longitudinal to
transverse $\gamma N$ cross-sections \cite{R1998} are used. Note
that such a procedure is equivalent to a fit to $(g_1)_{exp}$, but
it is more precise than the fit to the $g_1$ data themselves
actually presented by the experimental groups because here the
$g_1$ data are extracted in the same way for all of the data sets.
Note also, that in our analysis the logarithmic $Q^2$ dependence
of $h(x, Q^2)$ in Eq. (\ref{g1F2Rht}), which is not known in QCD,
is neglected. Compared to the principal $1/Q^2$ dependence it is
expected to be small and the accuracy of the present data does not
allow its determination. Therefore, the extracted from the data
values of $h(x)$ correspond to the mean $Q^2$ for each $x$-bean.

\subsection{Semi-inclusive DIS}

As in the inclusive DIS case, the measured double spin asymmetries
$A_{1N}^h$ for the polarized semi-inclusive deep inelastic
scattering, $\vec{l} + \vec{N} \rightarrow l + h +X$, where in
addition to the scattered lepton, hadron $h$ is also detected, can
be presented by the ratio of the spin structure functions
$g_{1N}^h$ and $g_{2N}^h$, and the unpolarized structure function
$F_{1N}^h$,
\begin{equation}
A_{1N}^h(x,z,Q^2)=\frac{g_{1N}^h(x,z,Q^2)-
\gamma^2g_{2N}^h(x,z,Q^2)}{F_{1N}^h(x,z,Q^2)}, \label{A1h}
\end{equation}
where $x$ is the Bjorken variable, $z=(p_h.p_N)/(p_N.q)$ is the
fractional energy of the hadrons in the the center-of-mass system
(c.m.s) frame of the nucleon and the virtual photon, and $q$ is
the usual notation for the photon four-momentum ($-q^2=Q^2$). In
(\ref{A1h}) the index N is used for the different targets and in
what follows it will be suppressed. Note also that in (\ref{A1h})
the contribution of the spin structure function $g_2^h$ to the
asymmetry $A_1^h$ can be neglected by two reasons. First, although
$g_2^h$ is not measured yet, it is expected to be small as in the
inclusive DIS case. Second, the $g_2^h$ term is multiplied by a
factor $\gamma^2=4M^2x^2/Q^2$ which in the kinematic region of the
present SIDIS experiments is negligible.

For the time being it is not known how to account for the HT and
TMC corrections in SIDIS processes. Fortunately, they should be
less important due to the kinematic region and the accuracy of the
present SIDIS data. So, in our QCD analysis we will use the
approximate equation:
\begin{equation}
A_{1N}^h(x,z,Q^2)=\frac{g_{1N}^h(x,z,Q^2)_{NLO}}{F_{1N}^h(x,z,Q^2)_{NLO}},
\label{g1hovF1h}
\end{equation}

In NLO QCD the structure functions $g_1^h$ and $F_1^h$ have the
following forms:
\begin{eqnarray}
 2g_1^h(x,z,Q^2)&=&\sum _{q,\bar{q}}
^{n_f}e_{q}^2\left\{\hspace{-0.4cm}\phantom{\int\limits_a^b}\Delta
q(x,Q^2)
D_q^h(z,Q^2)\right.\nonumber\\
&&+
\frac{\alpha_s(Q^2)}{2\pi}\left[\hspace{-0.4cm}\phantom{\int}\Delta
q\otimes \Delta C_{qq}^{(1)}
\otimes D_q^h\right.\nonumber\\
&&+ \Delta q\otimes \Delta C_{gq}^{(1)}\otimes D_g^h\nonumber\\
&&+ \left.\left.\hspace{-0.4cm}\phantom{\int}\Delta g\otimes
\Delta C_{qg}^{(1)}\otimes
D_q^h\right](x,z,Q^2)\phantom{\int\limits_a^b}\hspace{-0.4cm}\right\},
\label{g1h}
\end{eqnarray}
\begin{eqnarray}
2F_1^h(x,z,Q^2)&=&\sum _{q,\bar{q}}
^{n_f}e_{q}^2\left\{\hspace{-0.4cm}\phantom{\int\limits_a^b}q(x,Q^2)
D_q^h(z,Q^2)\right.\nonumber\\
&&+
\frac{\alpha_s(Q^2)}{2\pi}\left[\hspace{-0.4cm}\phantom{\int}q\otimes
C_{qq}^{(1)}\otimes D_q^h\right.\nonumber\\
&&+ q\otimes C_{gq}^{(1)}\otimes D_g^h\nonumber\\
&&+ \left.\left.\hspace{-0.4cm}\phantom{\int}g\otimes
C_{qg}^{(1)}\otimes
D_q^h\right](x,z,Q^2)\phantom{\int\limits_a^b}\hspace{-0.4cm}\right\}.
\label{F1h}
\end{eqnarray}
In Eqs. (\ref{g1h}) and (\ref{F1h}) $\Delta C_{ij}^{(1)}(x,z)$ and
$C_{ij}^{(1)}(x,z)$ are the NLO partonic coefficient functions in
the $\rm \overline{MS}$ scheme collected in \cite{FSV}.
$D_{q,\bar{q}}^h$, $D_g^h$ are the fragmentation functions (FFs)
for quarks, antiquarks and gluons, and $n_f$ is the number of
active flavors ($n_f=3$ in our present analysis).

\subsection{Method of analysis}

In our previous analyses of the inclusive DIS data the inverse
Mellin transformation method has been used to calculate the spin
structure function $g_1(x,Q^2)$ from its moments. The {\it double}
Mellin transform technique was developed by Stratmann and
Vogelsang and first applied in the NLO QCD analysis of SIDIS data
\cite{Double_Mellin}. We have used it to calculate the structure
functions $g_1^h(x,Q^2)$ and $F_1^h(x,Q^2)$ from their moments.
The expressions for the moments of the coefficient functions
$\Delta C_{ij}^{(1)}(x,z)$ and $C_{ij}^{(1)}(x,z)$ needed in these
calculations can been found in \cite{Double_Mellin}. For the
unpolarized parton densities we use the NLO MRST'02 PDFs
\cite{MRST02}, and for the fragmentation functions, the NLO DSS
set \cite{DSS} for pions, kaons and unindentified hadrons. The
main reason to use the MRST'02 set for the unpolarized PDFs is
that the DSS fragmentation functions were extracted from the data
using in the SIDIS sector the MRST'02 unpolarized PDFs and the
corresponding $\alpha_s(M^2_Z)$ value.

Compared with our previous fits to the inclusive DIS data only
(for example, see \cite{LSS07}) we use now a more general
parametrization for the input NLO polarized parton densities
at $Q^2_0 = 1~GeV^2$. It has the form for
$(\Delta u+\Delta \bar{u})$ and $(\Delta d+\Delta \bar{d})$
\begin{eqnarray}
\nonumber x(\Delta u+\Delta \bar{u})(x,Q^2_0)&=&A_{u+\bar{u}}x^
{\alpha_{u+\bar{u}}}
(1-x)^{\beta_{u+\bar{u}}}\\
\nonumber
&&(1+\epsilon_{u+\bar{u}}{\sqrt{x}}+
\gamma_{u+\bar{u}}x),\\[2mm]
\nonumber x(\Delta d+\Delta \bar{d})(x,Q^2_0)&=&A_{d+\bar{d}}x^
{\alpha_{d+\bar{d}}}
(1-x)^{\beta_{d+\bar{d}}}\\
&&(1+\epsilon_{d+\bar{d}}{\sqrt{x}}+
\gamma_{d+\bar{d}}x),
\label{input_ud}
\end{eqnarray}
and
\begin{eqnarray}
\nonumber
x\Delta \bar{u}(x,Q^2_0)&=&A_{\bar{u}}x^{\alpha_{\bar{u}}}
(1-x)^{\beta_{\bar{u}}}(1+\gamma_{\bar{u}}x),\\[2mm]
\nonumber
x\Delta \bar{d}(x,Q^2_0)&=&A_{\bar {d}}x^{\alpha_{\bar {d}}}
(1-x)^{\beta_{\bar {d}}}(1+\gamma_{\bar {d}}x),\\[2mm]
\nonumber
x\Delta \bar{s}(x,Q^2_0)&=&A_{\bar {s}}x^{\alpha_{\bar {s}}}
(1-x)^{\beta_{\bar {s}}}(1+\gamma_{\bar {s}}x),\\[2mm]
x\Delta G(x,Q^2_0)&=&A_{G}x^{\alpha_{G}}
(1-x)^{\beta_G}(1+\gamma_{G}x),
\label{inpu_sea_gl}
\end{eqnarray}
for the polarized sea quarks $\Delta {\bar{q}}$ and the gluon
parton densities. Since the accuracy of the present
SIDIS data is not enough to distinguish $\Delta s$ from $\Delta
\bar{s}$, we assume the relation $\Delta s(x)=\Delta \bar{s}(x)$.

As usual, the set of free parameters $\{a_i\}$ in (\ref{input_ud})
and (\ref{inpu_sea_gl}) is reduced by the well-known sum rules
\begin{equation}
a_3=g_{A}=\rm {F+D}=1.269~\pm~0.003,
\label{ga}
\end{equation}
\begin{equation}
a_8=3\rm {F-D}=0.585~\pm~0.025, \label{3FD}
\end{equation}
where $a_3$ and $a_8$ are nonsinglet combinations of the first
moments of the polarized parton densities corresponding to $3^{\rm
rd}$ and $8^{\rm th}$ components of the axial vector Cabibbo
current,
\begin{eqnarray}
a_3&=&(\Delta u+\Delta\bar{u})(Q^2) - (\Delta
d+\Delta\bar{d})(Q^2),\\[2mm]
\nonumber
a_8&=&(\Delta u +\Delta\bar{u})(Q^2) + (\Delta d +
\Delta\bar{d})(Q^2)\\[2mm] &&-2(\Delta s+\Delta\bar{s})(Q^2).
\end{eqnarray}

The constants $g_A$ in Eq. (\ref{ga}) and $a_8$ in Eq. (\ref{3FD})
are taken from \cite{PDG} and \cite{AAC00}, respectively. The sum
rule (\ref{ga}) reflects isospin SU(2) symmetry, whereas
(\ref{3FD}) is a consequence of the $SU(3)_f$ flavor symmetry
treatment of the hyperon $\beta$-decays. So, using the constraints
(\ref{ga}) and (\ref{3FD}) the parameters $A_{u+\bar{u}}$ and
$A_{d+\bar{d}}$ in (\ref{input_ud}) can be determined as functions
of the rest of the parameters connected with $(\Delta u+\Delta
\bar{u}),~(\Delta d+\Delta \bar{d})$ and $\Delta \bar{s}.$ In
addition, we assume that the parameters $\alpha_{u+\bar{u}}$ and
$\alpha_{d+\bar{d}}$ which control the small $x$ behavior of
$(\Delta u+\Delta \bar{u})$ and $(\Delta d+\Delta \bar{d})$ are
equal to those of $\Delta \bar{u}$ and $\Delta \bar{d}$,
respectively, which is reasonable as sea quarks likely dominate in
this region.

The large $x$ behavior of the polarized sea quarks and gluon
densities is mainly determined from the positivity constraints
\begin{equation}
\vert {\Delta f_i(x,Q^2_0)}\vert \leq f_i(x,Q^2_0),~~~~ \vert
{\Delta\bar{f_i}(x,Q^2_0)}\vert \leq \bar{f}_i(x,Q^2_0).
\label{pos}
\end{equation}

The constraints (\ref{pos}) are the consequence of a probabilistic
interpretation of the parton densities in the naive parton model,
which is still valid in LO QCD. Beyond LO the parton densities are
not physical quantities and the positivity constraints on the
polarized parton densities are more complicated. They follow from
the positivity condition for the polarized lepton-hadron
cross-sections $\Delta \sigma_i$ in terms of the unpolarized ones
($\vert {\Delta \sigma_i}\vert \leq \sigma_i$) and include also
the Wilson coefficient functions. It was shown \cite{AFR},
however, that for all practical purposes it is enough, at the
present stage, to consider LO positivity bounds for LO as well as
for NLO parton densities, since NLO corrections are only
relevant at the level of accuracy of a few percent.

For the unpolarized NLO parton densities on the RHS of
(\ref{pos}), we are using the MRST'02 parton densities
\cite{MRST02}. In order to guarantee fulfilling of the positivity
condition, we assume the following equation for the parameters
$\beta_i$ which control the large $x$ behavior of the polarized
sea quarks and gluons:
\begin{equation}
\beta_{\bar{q}}=\beta_G=\beta_{sea(MRST02)}=7.276.
\label{beta_i}
\end{equation}

The rest of parameters $\{a_i\}=\{A_i, \alpha_i, \beta_i,
\epsilon_i, \gamma_i\}$, as well as the unknown higher twist
corrections $h^N(x)$ to $g^N_1$ in (\ref{g1F2Rht}) have been
determined from a simultaneous fit to the DIS and SIDIS data. For
the determination of HT the measured $x$ region has been split
into 5 bins and to any $x$-bin two parameters $h_i^{(p)}$ and
$h_i^{(n)}$ have been attached \cite{LSS_HT}. For a deuteron
target the relation $h_i^{(d)}=0.925(h_i^{(p)}+ h_i^{(n)})/2~$ has
been used. So, to the set of parameters $\{a_i\}$ connected with
the input polarized PDFs (\ref{input_ud},~\ref{inpu_sea_gl}), 10
parameters for the HT corrections, $h_i^{(p)}$ and
$h_i^{(n)}~(i=1, 2, ..,5$), have been added.

In the polarized DIS and SIDIS processes the $Q^2$ range and the
accuracy of the data are much smaller than that in the unpolarized
case. That is why, in all calculations we have used a fixed value
of the NLO QCD parameter $~\Lambda_{\rm
\overline{MS}}(n_f=4)=311~{\rm MeV}$, which corresponds to
$~\alpha_s(M^2_{z}) = 0.1197$, as obtained by the MRST NLO QCD
analysis \cite{MRST02} of the world unpolarized data. The value of
$~\Lambda_{\rm \overline{MS}}$ above is slightly changed from that
of MRST'02 because the scale dependence of the strong running
coupling $\alpha_s(Q^2)$ is calculated using the so-called
"Denominator" representation \cite{alpha_Den}
\begin{equation}
\frac{\alpha_s(Q^2)}{4\pi}=\frac{1}{\beta_0ln(Q^2/\Lambda^2_{\rm
\overline{MS}})+\frac{\beta_1}{\beta_0}ln\{ln(Q^2/\Lambda^2_{\rm
\overline{MS}})+ \frac{\beta_1}{\beta^2_0}  \} },
\label{alphaDen}
\end{equation}
which is a more precise iterative solution of its renormalization
group equation at NLO accuracy. In (\ref{alphaDen}) $\beta_0=11-
2n_f/3,~\beta_1=102-38n_f/3$ and $~\Lambda_{\rm
\overline{MS}}(n_f=3,4,5)= 366,~311,~224~{\rm MeV}$. The number of
active flavors $n_f$ in $\alpha_s(Q^2)$ was fixed by the number of
quarks with $m^2_q\leq Q^2$ taking $m_c=1.43~GeV$ and
$m_b=4.3~GeV$.

The advantage of the analytic expression (\ref{alphaDen}) for
$\alpha_s$ is that: first, it reproduces with a very good accuracy
the numerical solution of the renormalization group equation needed
at small $Q^2$, down to $Q^2=1~GeV^2$, and second, for
$Q^2>4~GeV^2$ it practically coincides with the behavior of
$\alpha_s$ corresponding to its usual $1/ln(Q^2/\Lambda^2_{\rm
\overline{MS}})$-expansion at NLO \cite{PDG}.

\section{Results of analysis}

The numerical results of our global NLO QCD fit to the world
inclusive \cite{{EMC}, {SMC}, {COMPASSp}, {COMPASSd}, {SLAC142},
{SLAC143}, {SLAC154}, {SLAC155p}, {SLAC155d}, {HERMES}, {JLabn},
{CLAS06}} and semi-inclusive \cite{{HERMES}, {SMC_SI},
{COMPASSd_h}, {COMPASSd_pK}} DIS data are presented in Tables I
and II. The data used (841 experimental points for DIS and 202
experimental points for SIDIS) cover the following kinematic
regions: $\{0.005 \leq x \leq 0.75,~~1< Q^2 \leq 62~GeV^2\}$ for
DIS and $\{0.005 \leq x \leq 0.48,~~1< Q^2 \leq 60~GeV^2\}$ for
SIDIS processes.

In our analysis the minimization of the $\chi^2$ function is
performed using the program MINUIT at CERN \cite{MINUIT}. The
experimental errors are given by statistical and point-to-point
systematic errors added in quadrature. In the minimum of $\chi^2$
an accurate (a positive definite) error matrix is obtained and the
error bands of the polarized PDFs were calculated using the
standard Hessian method with $\Delta \chi^2 = 1$. We understand
that $\Delta \chi^2=1$ could be an underestimation of the
uncertainties of the polarized PDFs. If one wishes to use the
choice $\Delta \chi^2
> 1$, one has simply to scale our uncertainties of the polarized PDFs by
$(\Delta \chi^2)^{1/2}$. However, one has to keep in mind the
following points:

~~~(i) The systematic errors are partly correlated which would
lead to an overestimation of the errors when added in quadrature
with the statistical ones, which compensates a part of any
underestimation arising from using $\Delta \chi^2=1$.

~~(ii) In the analyses of the groups which present uncertainties
of polarized PDFs corresponding to $\Delta \chi^2>1$ (for an
example, $\Delta \chi^2={\rm UP} \sim ~N$, where $N$ is the number
of the free parameters), only the statistical errors are usually
taken into account.

~(iii) The different status of the PDFs and higher twist
parameters - practically they are not correlated.

(iv) The experimental errors in the polarized case are much larger
then those in the unpolarized one.
\begin{table}[h]
\caption{\label{tab:table1} Data used in our global NLO QCD
analysis, the individual $\chi^2$ for each set and the total
$\chi^2$ of the fit }
\begin{ruledtabular}
\begin{tabular}{ccccccc}
 Experiment &Process&$N_{data}$& $\chi^2$ \\ \hline
 EMC \cite{EMC}  &    DIS(p)   &  ~10 & ~4.2 \\
 SMC \cite{SMC}  &    DIS(p)   &  ~12 & ~5.5  \\
 SMC \cite{SMC}  &    DIS(d)   &  ~12 & 18.0   \\
 COMPASS \cite{COMPASSp} & DIS(p) &  ~15 & 12.0 \\
 COMPASS \cite{COMPASSd} & DIS(d) &  ~15 & ~8.4 \\
 SLAC/E142 \cite{SLAC142}& DIS(n) &  ~~8 & ~5.8 \\
 SLAC/E143 \cite{SLAC143}& DIS(p) &  ~28 & 17.8 \\
 SLAC/E143 \cite{SLAC143}& DIS(d) &  ~28 & 39.9 \\
 SLAC/E154 \cite{SLAC154}& DIS(n) &  ~11 & ~2.6 \\
 SLAC/E155 \cite{SLAC155p}& DIS(p) & ~24 & 25.5 \\
 SLAC/E155 \cite{SLAC155d}& DIS(d) & ~24 & 16.5 \\
 HERMES \cite{HERMES} & DIS(p)  & ~~9 & ~5.4 \\
 HERMES \cite{HERMES} & DIS(d)  & ~~9 & ~6.8 \\
 JLab-Hall A \cite{JLabn}& DIS(n)& ~~3& ~0.3 \\
 CLAS \cite{CLAS06} &  DIS(p) & 151 &   119.9 \\
 CLAS \cite{CLAS06} &  DIS(d) & 482 &   427.9 \\
\hline
 SMC \cite{SMC_SI} & SIDIS(p,$h^{+}$)& ~12 & 18.1\\
 SMC \cite{SMC_SI} & SIDIS(p,$h^{-}$)& ~12 & 11.2\\
 SMC \cite{SMC_SI} & SIDIS(d,$h^{+}$)& ~12 & ~9.4\\
 SMC \cite{SMC_SI} & SIDIS(d,$h^{-}$)& ~12 & 13.8\\
HERMES \cite{HERMES}&SIDIS(p,$h^{+}$)& ~~9 & ~5.9\\
HERMES \cite{HERMES}&SIDIS(p,$h^{-}$)& ~~9 & ~5.3\\
HERMES \cite{HERMES}&SIDIS(d,$h^{+}$)& ~~9 & 10.5\\
HERMES \cite{HERMES}&SIDIS(d,$h^{-}$)& ~~9 & ~4.8\\
HERMES \cite{HERMES}&SIDIS(p,$\pi^{+}$)& ~~9 & ~9.9\\
HERMES \cite{HERMES}&SIDIS(p,$\pi^{-}$)& ~~9 & ~5.1\\
HERMES \cite{HERMES}&SIDIS(d,$\pi^{+}$)& ~~9 & ~8.6\\
HERMES \cite{HERMES}&SIDIS(d,$\pi^{-}$)& ~~9 & 19.8\\
HERMES \cite{HERMES}&SIDIS(d,$K^{+}$)& ~~9 & ~6.7\\
HERMES \cite{HERMES}&SIDIS(d,$K^{-}$)& ~~9 & ~5.6\\
COMPASS \cite{COMPASSd_h}& SIDIS(d,$h^{+}$)& ~12 & ~7.6\\
COMPASS \cite{COMPASSd_h}& SIDIS(d,$h^{-}$)& ~12 & 10.9\\
COMPASS \cite{COMPASSd_pK}& SIDIS(d,$\pi^{+}$)& ~10 & ~2.6\\
COMPASS \cite{COMPASSd_pK}& SIDIS(d,$\pi^{-}$)& ~10 & ~4.5\\
COMPASS \cite{COMPASSd_pK}& SIDIS(d,$K^{+}$)& ~10 & ~7.8\\
COMPASS \cite{COMPASSd_pK}& SIDIS(d,$K^{-}$)& ~10 & 13.7\\
\hline
{\bf \ TOTAL}: &        &   1043 &   898.6  \\
\end{tabular}
\end{ruledtabular}
\end{table}

In Table I the data sets, both for inclusive and semi-inclusive
DIS, used in our analysis are listed and the corresponding values
of $\chi^2$ obtained from the best fit to the data are presented.
As seen from Table I, a good description of the data is achieved
for both the inclusive ($\chi^2_{NrP}$=0.85) and semi-inclusive
($\chi^2_{NrP}$=0.90) processes (NrP is the number of
corresponding experimental points). The total value of
$\chi^2_{DF}$ is 0.88. The quality of the fit to the data is
demonstrated in Fig. 1 for some of the SIDIS asymmetries obtained
by the HERMES and COMPASS Collaborations.
\begin{figure}[ht]
\includegraphics[scale=1.10]{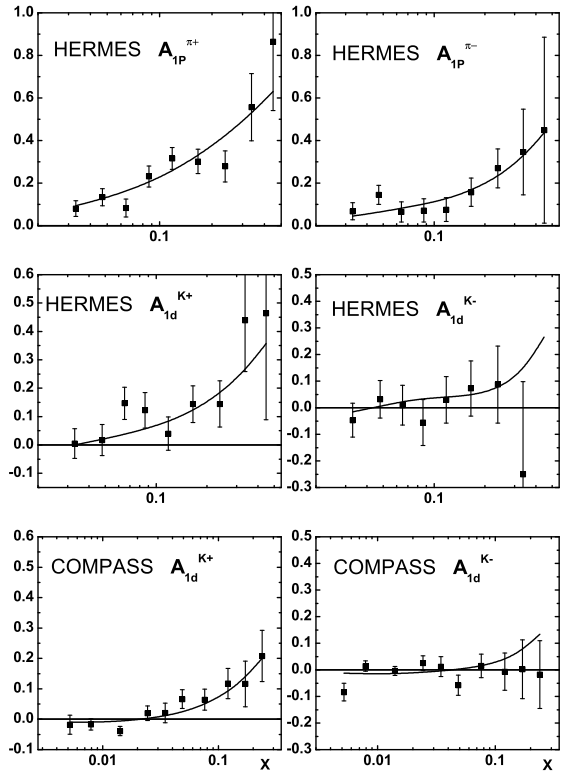}
 \caption{Comparison of our NLO QCD results for the SIDIS asymmetries
with the data at measured $x$ and $Q^2$.
 \label{data_h}}
\end{figure}

The values of the parameters attached to the input polarized PDFs
obtained from the best fit to the data are presented in Table II.
The errors correspond to $\Delta \chi^2=1$. Note also that only
the experimental errors (statistical and systematic) are taken
into account in their calculation. It was impossible to determine
from the fit the parameters $\epsilon_{d+\bar{d}}$ and
$\gamma_{\bar{d}}$ in Eqs. (\ref{input_ud}) and
(\ref{inpu_sea_gl}), respectively, so they were eliminated i.e.
put equal to zero. Note that the central value of
$\gamma_{\bar{d}}$ obtained from the fits was always positive. So
that its elimination does not change the negative behavior of
$x\Delta \bar{d}(x)$ for any $x$ in the measured region.
\begin{table*}
\caption{\label{tab:table2} The parameters of the NLO($\rm
\overline{MS}$) input polarized PDFs at $Q^2=1~\rm GeV^2$ obtained
from the best fit to the data. The parameters marked by (*)
are fixed. }
\begin{ruledtabular}
\begin{tabular}{ccccccc}
  flavor &  A  &  $\alpha$ &  $\beta$ & $\epsilon$ & $\gamma$  \\ \hline
 $u+\bar{u}$& ~1.097$^*$ & 0.782~$\pm$~0.165 & 3.335~$\pm$~0.154 &
-1.779~$\pm$~0.896 & 10.20~$\pm$~5.61~ \\
 $d+\bar{d}$& -0.394$^*$ & 0.547~$\pm$~0.118 & 4.056~$\pm$~0.543 &
0 & 6.758~$\pm$~5.366  \\
$\bar{u}$& 0.334~$\pm$~0.174 & 0.782~$\pm$~0.165 & 7.267$^*$ &
0 & -5.136~$\pm$~2.414 \\
$\bar{d}$& -0.133~$\pm$~0.075 & 0.547~$\pm$~0.118 & 7.267$^*$ &
0 & 0 \\
$\bar{s}$ & -0.00352~$\pm$~0.00194 & 0.0458~$\pm$~0.0206 & 7.267$^*$ &
0 & -39.02~$\pm$~29.62 \\
G & -68.23~$\pm$~65.79 & 1.975~$\pm$~0.459 & 7.267$^*$ & 0 &
-3.536~$\pm$~1.089 \\
\end{tabular}
\end{ruledtabular}
\end{table*}

\subsection{The role of semi-inclusive DIS data in determining the
polarized sea-quark densities}

Let us discuss the impact of semi-inclusive DIS data on the
polarized PDFs. Because of SIDIS data a flavor decomposition of
the polarized sea is achieved and the light antiquark polarized
densities $\Delta \bar{u}(x)$ and $\Delta \bar{d}(x)$ are
determined without any additional assumptions. While $\Delta
\bar{d}(x)$ is negative for any $x$ in the measured $x$ region,
$\Delta \bar{u}(x)$ is a positive, passes zero around $x=0.2$ and
becomes negative for large $x$. Sign-changing solutions are also
found for the polarized strange sea $\Delta \bar{s}(x)$ and gluon
$\Delta G(x)$ densities. The sign-changing behavior for $\Delta
G(x)$ is not surprising since it was already found from the
analysis of the inclusive DIS data alone \cite{LSS07}. On the
other hand, on the basis of results from all published analyses of
inclusive DIS, we consider the sign-changing solution for $\Delta
\bar{s}(x)$ quite puzzling. The central values of the sea-quark
and gluon polarized densities together with their error bands are
presented and compared to those of DSSV (dashed curves) in Fig. 2.
\begin{figure}[h]
\includegraphics[scale=0.80]{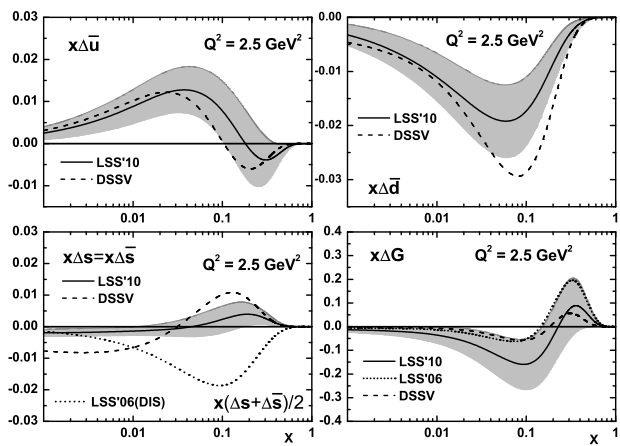}
 \caption{Our NLO sea quarks and gluon polarized PDFs at
$Q^2=2.5~GeV^2$ in the $\rm \overline{MS}$ scheme. For comparison
the DSSV PDFs \cite{DSSV} are also presented.
\label{seaPDFs}}
\end{figure}
The main difference between the Leader-Sidorov-Stamenov (LSS) and
DSSV sets is in the strange sea-quark density $\Delta \bar
{s}(x)$. Although the first moments are almost equal (-0.054 and
-0.055 at $Q^2=1~GeV^2$ for LSS and DSSV, respectively), our
$\Delta \bar {s}(x)$ is less negative for $x<0.03$ and less
positive for $x>0.03$. Note that DSSV have used an additional
constraint $\alpha_{\bar{s}}=\alpha_{\bar{d}}$ for the parameters
$\alpha_{\bar{s}}$ and $\alpha_{\bar{d}}$ which means a similar
small $x$ behavior for the sea-quark densities $\Delta \bar
{s}(x)$ and$\Delta \bar {d}(x)$. We do not think this assumption
is reasonable. The central values of our gluon density and its
first moment are rather different from those of DSSV, however they
coincide within the errors which are still large in the measured
$x$ region.

In Fig. 2 our LSS'06 PDFs (dotted curves) \cite{LSS07} obtained
from the NLO QCD analysis of the world inclusive DIS data are
presented too. While the light antiquark polarized densities
$\Delta \bar{u}(x)$ and $\Delta \bar{d}(x)$ cannot be, in
principle, determined from polarized inclusive DIS data, the sum
$(\Delta s +\Delta \bar{s})(x, Q^2)$ \emph{is} well determined and
all the NLO QCD analyses yield for this sum a {\it negative} value
for any $x$ in the measured region (for example, see Refs.
\cite{{groups}, {LSS07}, {groups_add}}). In these analyses,
however, a term like $(1+\gamma x)$, which would permit a
sign-change, was not included  in the input parametrization of
$(\Delta s +\Delta \bar{s})(x, Q^2_0)/2$ \cite{note1}. We
therefore re-analyzed the world polarized inclusive DIS data using
such a term in the input strange sea-quark density
\begin{equation}
(\Delta s +\Delta \bar{s})(x, Q^2_0)/2 =
Ax^{\alpha}
(1-x)^{\beta}(1+\gamma x).
\end{equation}

Our preliminary results confirm the previous ones, namely, that
$(\Delta s +\Delta \bar{s})(x, Q^2)/2$ is negative in the measured
$(x,Q^2)$ region. So, the behavior of the polarized strange quark
density remains controversial. Note that in the presence of SIDIS
data $\Delta s$ and $\Delta \bar{s}$ can, in principle, be
separately determined, as was done recently by the COMPASS
Collaboration, where it was shown \cite{COMPASS_dels} that there
is no significant difference between $\Delta s(x)$ and $\Delta
\bar {s}(x)$ in the $x$-range covered by their inclusive and
semi-inclusive DIS data. This result was obtained in the LO QCD
approximation, with the additional assumption that the SIDIS
asymmetries are $Q^2$-independent. We checked the latter
assumption using in the calculations of the asymmetries our NLO
PDFs,  and found it not quite correct. Also, the errors of the
extracted values of the difference $x(\Delta s(x)-\Delta \bar
{s}(x))$ are rather too large to allow us to conclude that the
assumption $\Delta s(x)=\Delta \bar{s}(x)$ used in our analysis
and that of the DSSV is correct. So, if it is not correct, it
might possibly be the cause  that $(\Delta s +\Delta \bar{s})(x,
Q^2)/2$ densities obtained from the analyses of inclusive DIS data
and combined inclusive and semi-inclusive DIS data sets,
respectively, are in contradiction. However, at first glance, it
looks as if the difference between $\Delta s $ and $\Delta
\bar{s}$ would have to be quite significant and might contradict
the COMPASS results. Perhaps a more important issue is the
sensitivity of the results to the form of the fragmentation
functions. An analysis by the COMPASS group  \cite{COMPASSd_pK}
demonstrated that the determination of $\Delta \bar{s}(x)$
strongly depends on the set of the fragmentation functions used in
the analysis and that the DSS FFs are crucially responsible for
the unexpected behavior of $\Delta \bar{s}(x)$ obtained from the
combined analysis. The study of the sensitivity of $\Delta
\bar{s}(x)$ to different sets of FFs used in the analysis is one
of the key points we plan to investigate in the future.

In Fig. 3 we present our results for the polarized $\Delta u(x)$
and $\Delta d(x)$ densities at $Q^2=2.5~GeV^2$, which are
consistent with those obtained by DSSV (dashed curves). As
expected, the SIDIS data do not influence essentially the sums
$(\Delta u(x)+\Delta \bar{u}(x))$ and $(\Delta d(x)+\Delta
\bar{d}(x))$ already well determined from the analysis of the
inclusive DIS data. This fact is illustrated in Fig. 3 where our
results from the combined analysis are compared with our LSS'06
PDFs.
\begin{figure}[t]
\includegraphics[scale=0.80]{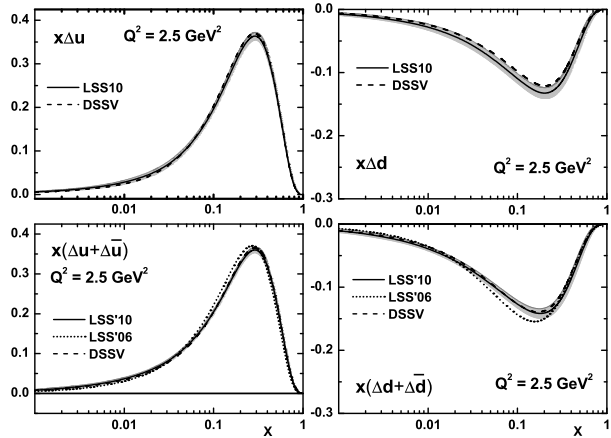}
 \caption{Our results for $\Delta u$, $\Delta d$,
$(\Delta u+\Delta \bar{u})$ and $(\Delta d+\Delta \bar{d})$
polarized parton densities at NLO approximation. DSSV \cite{DSSV}
as well as LSS'06 \cite{LSS07} results for the corresponding
densities are presented too.
\label{du_dd}}
\end{figure}

\subsection{High twist effects}

As mentioned in Sec. II A, in contrast to other combined analyses
of the inclusive and semi-inclusive DIS data, we take into account
the target mass and higher twist corrections in the DIS sector.
The values of the HT corrections to $g_1$ extracted from the data
in this analysis are presented in Table III and illustrated in
Fig.4.
\begin{table}
\caption{\label{tab:table3} The values of higher twist corrections
extracted from the data in a model independent way. }
\begin{ruledtabular}
\begin{tabular}{ccccccc}
 $x_i$ &~~~ $h^p(x_i)~[GeV^2]$ &~~~~~~ $x_i$&~~~ $h^n(x_i)~[GeV^2]$\\ \hline
 0.028 &~~~ -0.048~$\pm$~0.037 &~~~~~ 0.028 &~~~ 0.093~$\pm$~0.051 \\
 0.100 &~~~ -0.084~$\pm$~0.016 &~~~~~ 0.100 &~~~ 0.041~$\pm$~0.036 \\
 0.200 &~~~ -0.053~$\pm$~0.012 &~~~~~ 0.200 &~~~ 0.000~$\pm$~0.027 \\
 0.350 &~~~ -0.045~$\pm$~0.011 &~~~~~ 0.325 &~~~ 0.022~$\pm$~0.018 \\
 0.600 &~~~ -0.012~$\pm$~0.014 &~~~~~ 0.500 &~~~ 0.017~$\pm$~0.015 \\
\end{tabular}
\end{ruledtabular}
\end{table}

Compared to the HT(LSS'06) corrections obtained in our analysis of
the inclusive DIS data alone \cite{LSS07} the values of the HT
corrections for the proton target are practically not changed,
while the central values of HT corrections for the neutron target
are smaller in the region $x<0.2$, but in agreement with $\rm
HT^{(n)}$(LSS'06) within the errors, excepting the $x$ region
around $x=0.1$. We consider the tendency of the $\rm HT^{(n)}$
corrections to be smaller in the region $x<0.2$ to be  a result of
the new behavior of $\Delta s(x)$, i.e. positive for $x > 0.03$.
The positive contribution in $g_1^n$ from $\Delta s(x)$ should be
compensated by a less positive $\rm HT^{(n)}$ contribution in this
region. Since the biggest difference between the values of $\Delta
s(x)_{\rm (DIS+SIDIS)}$ and $\Delta s(x)_{\rm DIS}$ is in the
region $x\sim 0.1$ (see Fig. 2), this effect is biggest in this
$x$ region.
\begin{figure}[t]
\includegraphics[scale=0.70]{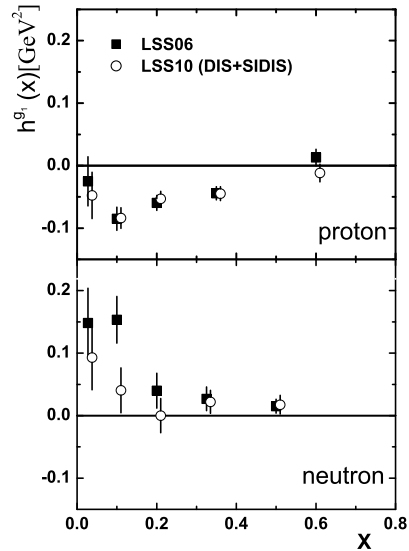}
 \caption{Comparison between HT values corresponding to the fits
of inclusive DIS \cite{LSS07} and combined inclusive and SIDIS
data set (this analysis). \label{HT}}
\end{figure}
The impact of $\Delta s(x)$ on HT corrections is visible mainly
for the neutron target because the contribution of $\Delta s(x)$
in $g_1^n$ is relatively larger than that in $g_1^p$.

Let us briefly discuss the values of the first moments of
the higher twist corrections to the proton and neutron structure
function $g_1$. Using the values for $h^N(x)$ from Table III we obtain
for their first moments in the experimental region:
\begin{equation}
\bar h^N\equiv\int^{0.75}_{0.0045}h^N(x)dx,~~~~(N=p, n)
\label{momHT}
\end{equation}
$\bar h^p= (- 0.028\pm 0.005)~\rm GeV^2$ for the proton and $\bar
h^n=(0.018 \pm 0.008)~\rm GeV^2$ for the neutron target. As a
result, for the nonsinglet $(\bar h^p - \bar h^n)$ and the singlet
$(\bar h^p + \bar h^n)$ we obtain $(- 0.046\pm 0.009)~\rm GeV^2$
and $(-0.011\pm 0.009)~\rm GeV^2$, respectively. The statistical
and systematic errors are added in quadrature. Note that in our
notation $h=\int^{1}_{0}h(x)dx=4M^2(d_2 + f_2)/9$, where $d_2$ and
$f_2$ are the well known quantities, connected with the matrix
elements of twist 3 and twist 4 operators, respectively
\cite{d2f2}, and $M$ is the mass of the nucleon.

Our value for the nonsinglet $(\bar h^p - \bar h^n)$ is well
consistent within the errors with those extracted directly from
the recent analyses of the first moment of the nonsinglet spin
structure function $g_1^{(p-n)}(x, Q^2)$ \cite{{h_nons_Deur},
{h_nons_Dubna}}. Note that our value for the nonsinglet $(\bar h^p
- \bar h^n)$ is also in agreement with the QCD sum rule estimates
\cite{Balitsky:1990jb} as well as with the instanton model
predictions \cite{Balla:1997hf,SidWeiss}. The values obtained for
the nonsinglet $(\bar h^p - \bar h^n)$ and singlet $(\bar h^p +
\bar h^n)$ quantities are in qualitative agreement with the
relation $|h^p + h^n| << |h^p - h^n|$ derived in the large $\rm
N_c$ limit in QCD \cite{Balla:1997hf}.

Our results on the higher twist effects are not in agreement
with those obtained in \cite{BB2010}, where the authors
find no evidence for HT.

\subsection{The sign of the gluon polarization}

We have found that the combined NLO QCD analysis of the present
polarized inclusive DIS and SIDIS data cannot rule out the
solution with a positive gluon polarization. The values of
$\chi^2/DF$ corresponding to the fits with sign-changing $x\Delta
G(x,Q^2)$ and positive $x\Delta G(x,Q^2)$ are practically the
same: $\chi^2/DF({\rm node}~x\Delta G )=0.883$ and
$\chi^2/DF(x\Delta G >0)=0.888$, and the data cannot distinguish
between these two solutions (see Fig. 5).
\begin{figure}[h]
\includegraphics[scale=0.60]{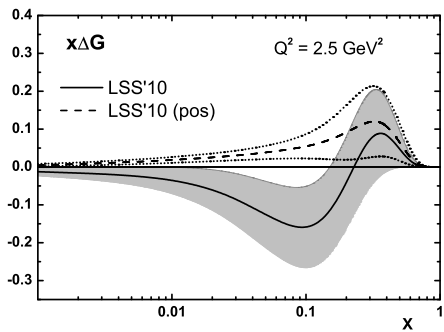}
 \caption{Comparison between the positive and sign-changing
 gluon densities. The corresponding error bands are also shown.
 \label{posdelG}}
\end{figure}

The corresponding sea-quark densities are shown in Fig. 6. As
seen, the sea-quark densities obtained in the fits with positive
and sign-changing $x\Delta G(x)$ are almost identical. Note that
the extracted HT values corresponding to both fits are also
effectively identical. As a result, one can conclude that
including the SIDIS data in the QCD analysis does not help to
constrain better the polarized gluon density.
\begin{figure}[h]
\includegraphics[scale=0.80]{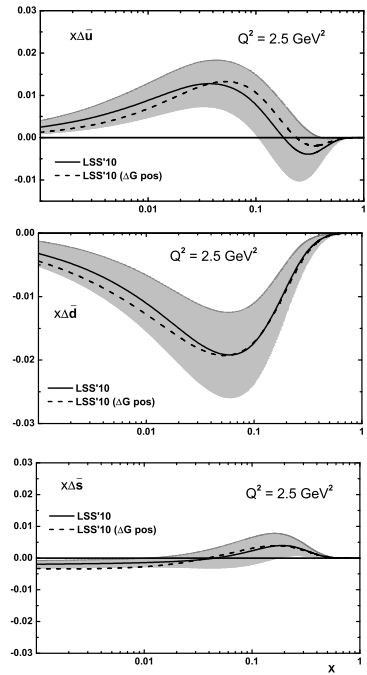}
 \caption{Comparison between the sea-quark densities corresponding
 to positive and changing in sign gluon densities.
\label{sea_dGpos}}
\end{figure}

In Fig. 7 the ratio $\Delta G(x)/G(x)$ calculated for both the
sign-changing and positive solutions for $\Delta G(x)$ obtained in
our NLO QCD analysis is compared with the directly measured values
of $\Delta G/G$ obtained from a quasireal photoproduction of high
$p_t$ hadron pairs \cite{{SMC_high_pt}, {COMPASS_high_pt},
{HERMES_high_pt}}, and from the open charm production
\cite{open_charm} measurements. For the unpolarized gluon density
$G(x)$ in the ratio above we have used that of the NLO MRST'02
\cite{MRST02}. The theoretical curves are given for $\mu^2=3~\rm
GeV^2$ (high $p_t$ hadron pairs) and  $\mu^2=13~\rm GeV^2$ (open
charm). As seen from Fig. 7,  both solutions for the polarized
gluon density are well consistent with the experimental values of
$\Delta G/G$. It should be noted, however, that in the extraction
of $\Delta G/G$ by the experiments a LO QCD treatment has been
used. A NLO extraction of the measured values is needed in order
for this comparison to be quite correct. In conclusion, the
magnitude of the gluon density $x\Delta G(x)$ obtained from our
combined NLO QCD analysis of inclusive and semi-inclusive DIS data
and independently, from the photon-gluon fusion processes, is
small in the region $x\simeq 0.08-0.2$.
\begin{figure}[h]
\includegraphics[scale=0.58]{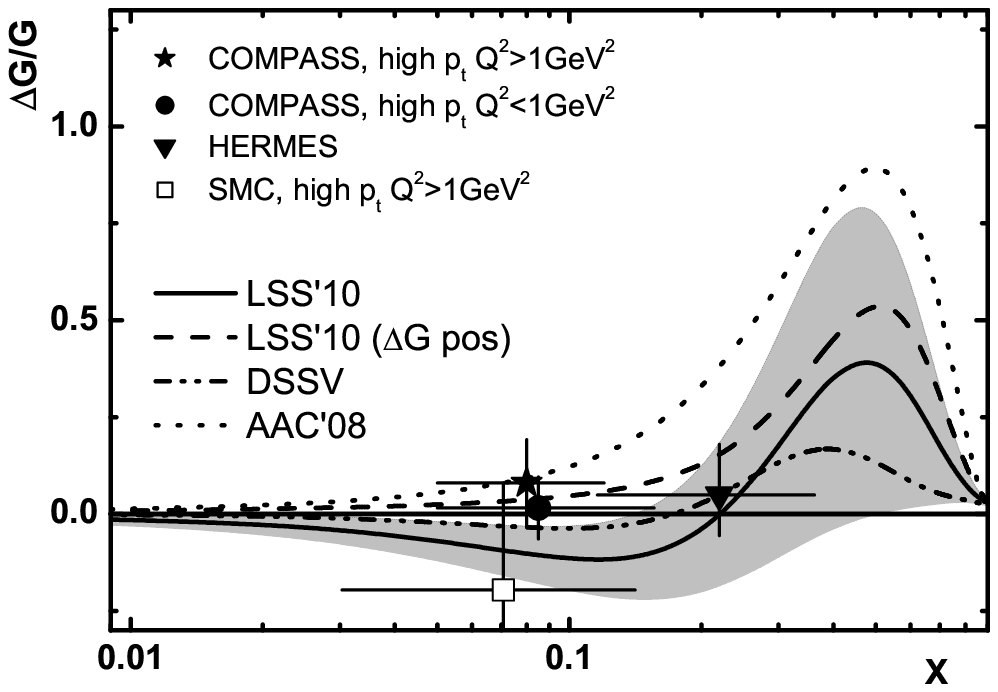}
\includegraphics[scale=0.65]{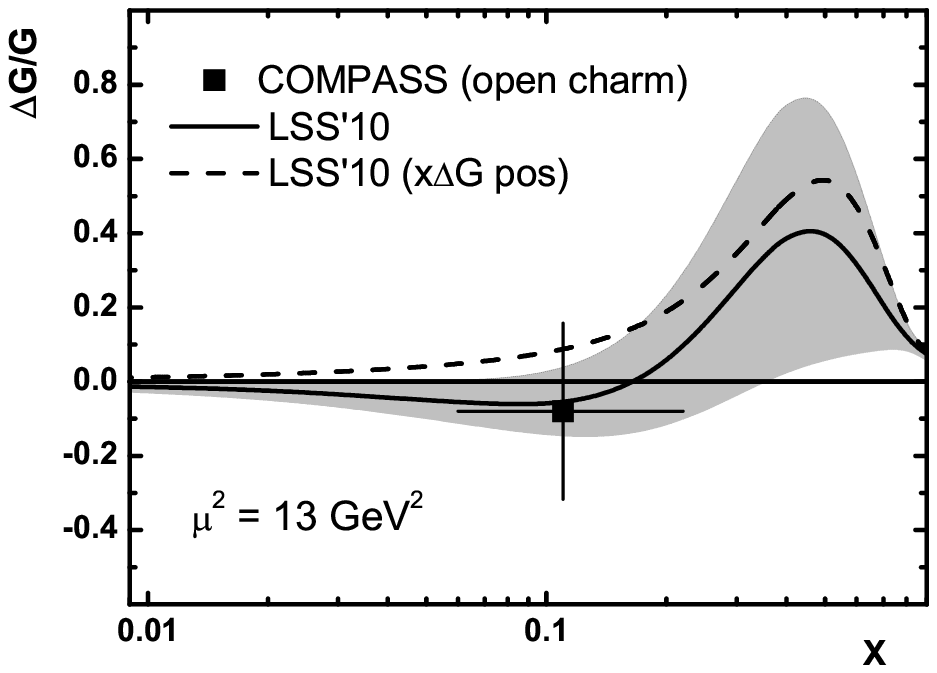}
 \caption{Comparison between the experimental data and
NLO($\rm \overline{MS}$) curves for the ratio $\Delta G(x)/G(x)$
at $Q^2=3~\rm GeV^2$ ({\bf top} - high $p_t$ pairs) and
$Q^2=13~\rm GeV^2$ ({\bf bottom} - open charm) corresponding to
positive and sign-changing $x\Delta G$. Error bars represent the
total (statistical and systematic) errors. The horizontal bar on
each point shows the $x$-range of the measurement. The NLO AAC
(second listing of Ref. in \cite{groups}) and DSSV \cite{DSSV}
curves on $\Delta G(x)/G(x)$ are also presented. \label{delGovG} }
\end{figure}

When this analysis was finished, the COMPASS Collaboration
reported the first data on the asymmetries
$A_{1,p}^{\pi_{+(-)}},~A_{1,p}^{K_{+(-)}}$ for charged pions and
kaons produced on a proton target \cite{COMPASS_dels}. As seen in
Fig. 8, our predictions for these asymmetries are in very good
agreement with the data at measured $x$ and $Q^2$.
\begin{figure}[h]
\includegraphics[scale=0.80]{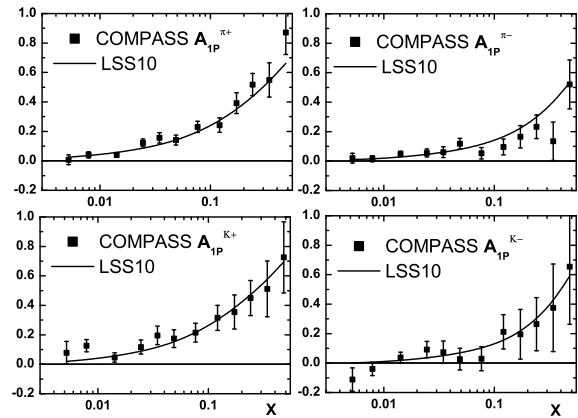}
 \caption{Our predictions for the COMPASS asymmetries for charged
 pions and kaons produced on a proton target.
 \label{LSS10_pred}}
\end{figure}

\subsection{The spin sum rule}

Let us finally discuss the present status of the proton spin sum
rule. Using the values for $\Delta \Sigma(Q^2)$ and $\Delta
G(Q^2)$ at $Q^2=4~GeV^2$, the first moments of the quark singlet
$\Delta \Sigma(x,Q^2)$ and gluon $\Delta G(x, Q^2)$ densities,
obtained in our analysis (see Table IV) one  finds for the spin of
the proton:
\begin{table*}
\caption{\label{tab:table4} First moments of polarized PDFs at
$Q^2=4~GeV^2$. The corresponding DSSV values are also presented.   }
\begin{ruledtabular}
\begin{tabular}{ccccccc}
Fit & $\Delta \bar{s}$ & $\Delta G$ &
$\Delta \Sigma$ \\ \hline
 LSS'10 (pos $x\Delta G$)  & -0.063~$\pm$~0.004 & ~0.316~$\pm$~0.190& 0.207~$\pm$~0.034 \\
 LSS'10 (node $x\Delta G$) & -0.055~$\pm$~0.006 &-0.339~$\pm$~0.458& 0.254~$\pm$~0.042 \\
 DSSV (node $x\Delta G$) & -0.056 & -0.096 & 0.245 \\
\end{tabular}
\end{ruledtabular}
\end{table*}
\begin{eqnarray}
J_z = \frac{1}{2}&=&\frac{1}{2}\Delta \Sigma(Q^2)+\Delta G(Q^2)+L_z(Q^2)
\nonumber\\
&=&-0.21 \pm 0.46 + L_z(Q^2)~~ ({\rm node}~\Delta G), \nonumber\\
&=&~~0.42 \pm 0.19 + L_z(Q^2)~~ ({\rm pos}~\Delta G).
\label{SSR}
\end{eqnarray}
In Eq. (\ref{SSR}) $L_z(Q^2)$ is the sum of the angular orbital
momenta of the quarks and gluons. Although the central values of
the quark-gluon contribution in (\ref{SSR}) are very different in
the two cases, in view of the large uncertainty coming mainly from
the gluons, one cannot  yet come to a definite conclusion about
the contribution of the orbital angular momentum to the total spin
of the proton.

\section{Summary}

A new combined NLO QCD analysis of the polarized inclusive and
semi-inclusive DIS data is presented. In contrast to previous
combined analyses, the $1/Q^2$ terms (kinematic - target mass
corrections, and dynamic - higher twist corrections) to the
nucleon spin structure function $g_1$ are taken into account. The
new results for the  PDFs are compared to both the LSS'06 PDFs
obtained from a fit to the inclusive DIS data alone, and to those
obtained from the DSSV global analysis. The role of the
semi-inclusive data in determining the polarized sea quarks is
discussed. Because of SIDIS data $\Delta \bar{u}(x,Q^2)$ and
$\Delta \bar{d}(x,Q^2)$, as well $\Delta u(x,Q^2)$ and $\Delta
d(x,Q^2)$ are determined without additional assumptions about the
light sea quarks. The SIDIS data, analyzed under the assumption $
\Delta s(x,Q^2)=   \Delta \bar{s}(x,Q^2)$, imposes a sign-changing
$\Delta \bar{s}(x,Q^2)$, as in the DSSV analysis, but our values
are smaller in magnitude, less negative at $x < 0.03$ and less
positive for $x > 0.03$. Note that $\Delta \bar{s}(x,Q^2)_{\rm
SIDIS}$ differs essentially from the negative $\frac{1}{2}(\Delta
s +\Delta \bar{s})(x, Q^2)_{\rm DIS}$ obtained from all the QCD
analyses of inclusive DIS data. As was mentioned above the
behavior of $\Delta \bar{s}(x,Q^2)_{\rm SIDIS}$ strongly depends
on the  fragmentation functions used in our analysis and that of
the DSSV. A further detailed analysis of the sensitivity of
$\Delta \bar{s}(x,Q^2)$ to the FFs is needed, and any model
independent constraints on FFs would help. Another possible reason
for this disagreement could be the assumption $\Delta s(x,Q_0^2)
=\Delta \bar{s}(x,Q_0^2)$ made in the global analyses. However,
this would probably require a significant difference between
$\Delta s $ and $\Delta \bar{s}$. In any case, obtaining a final
and unequivocal result for $\Delta \bar{s}(x)$ remains a challenge
for further research on the internal spin structure of the
nucleon.

We have found also that the polarized gluon density is still
ambiguous, and the present polarized DIS and SIDIS data cannot
distinguish between the positive and sign-changing gluon densities
$\Delta G(x)$. This ambiguity is the main reason that the
quark-gluon contribution into the total spin of the proton is
still not well determined.

Finally, our combined NLO QCD analysis confirms our previous
results on the higher twist corrections to the nucleon spin
structure function $g_1^N$, namely, that they are not negligible
in the preasymptotic region and have to be accounted for in order
to extract correctly the polarized PDFs.

\begin{acknowledgments}
This research was supported by the JINR-Bulgaria Collaborative
Grant, by the RFBR Grants (No. 08-01-00686 and No. 09-02-01149)
and by the Bulgarian National Science Foundation under Contract
No. 288/2008.
\end{acknowledgments}


\begin{thebibliography}{99}
\bibitem{note1}
Note that in all the analyses \cite{{groups}, {LSS07}, {groups_add}}
the convention $\Delta s(x) = \Delta \bar{s}(x)$ has been used, but
one has to remember that only their sum can be extracted from the
inclusive DIS data.
\end{thebibliography}
\end{document}